\documentclass[conference]{IEEEtran}
\IEEEoverridecommandlockouts
\usepackage{cite}
\usepackage{amsmath,amssymb,amsfonts}
\usepackage{algorithmic}
\usepackage{graphicx}
\graphicspath{ {./image/} }
\usepackage{textcomp}
\usepackage{xcolor}
\usepackage{tikz}
\usepackage{enumitem}
\usepackage{flexisym}
\usepackage{multirow}
\usepackage{color, colortbl}
\definecolor{maroon}{cmyk}{0,0.87,0.68,0.32}
\def\BibTeX{{\rm B\kern-.05em{\sc i\kern-.025em b}\kern-.08em
    T\kern-.1667em\lower.7ex\hbox{E}\kern-.125emX}}
\makeatletter

\def\ps@IEEEtitlepagestyle{%
  \def\@oddfoot{\mycopyrightnotice}%
  \def\@evenfoot{}%
}
\def\mycopyrightnotice{%
  {\footnotesize  978-1-7281-5409-1/20/\$31.00\ \textsuperscript{\textcopyright}2020 IEEE\hfill}
  \gdef\mycopyrightnotice{}
}

\newcommand*{\rom}[1]{\expandafter\@slowromancap\romannumeral #1@}
\makeatother
\newcommand\solidcircle{\tikz\draw[black,fill=black] (0,0) circle (.5ex);}
\usepackage{colortbl}
\usetikzlibrary{calc}
\usepackage{zref-savepos}
\newcounter{NoTableEntry}
\renewcommand*{\theNoTableEntry}{NTE-\the\value{NoTableEntry}}

\newcommand*{\notableentry}{%
  \multicolumn{1}{@{}c@{}|}{%
    \stepcounter{NoTableEntry}%
    \vadjust pre{\zsavepos{\theNoTableEntry t}}
    \vadjust{\zsavepos{\theNoTableEntry b}}
    \zsavepos{\theNoTableEntry l}
    \hspace{0pt plus 1filll}%
    \zsavepos{\theNoTableEntry r}
    \tikz[overlay]{%
      \draw[black]
        let
          \n{llx}={\zposx{\theNoTableEntry l}sp-\zposx{\theNoTableEntry r}sp},
          \n{urx}={0},
          \n{lly}={\zposy{\theNoTableEntry b}sp-\zposy{\theNoTableEntry r}sp},
          \n{ury}={\zposy{\theNoTableEntry t}sp-\zposy{\theNoTableEntry r}sp}
        in
        (\n{llx}, \n{lly}) -- (\n{urx}, \n{ury})
        (\n{llx}, \n{ury}) -- (\n{urx}, \n{lly})
      ;
    }%
  }%
}

\usepackage{amsmath}

\usepackage{textcomp}
\begin{document}
\bstctlcite{IEEEexample:BSTcontrol}
\setlength{\textfloatsep}{0pt}
\setlength{\intextsep}{2pt}
\setlength{\belowcaptionskip}{2pt}
\setlength{\abovedisplayskip}{1pt}
\setlength{\belowdisplayskip}{1pt}
\setlength\abovedisplayshortskip{1pt}
\setlength\belowdisplayshortskip{1pt}
\title{ Early RTL Analysis for SCA Vulnerability in Fuzzy Extractors of Memory-Based PUF Enabled Devices}
\author{Xinhui Lai$^1$, Maksim Jenihhin$^1$,  Georgios Selimis$^2$, Sven Goossens$^2$, Roel Maes$^2$, Kolin Paul$^{3}$ \\
$^1$ Department of Computer Systems, Tallinn University of Technology, Estonia\\
$^2$ Intrinsic ID, The Netherlands\\
$^3$ Department of Computer Science \& Engg, Indian Institute of Technology Delhi, India \\
Email: xinhui.lai@taltech.ee
}

\onecolumn
\noindent\textcopyright 2020 IEEE. Personal use of this material is permitted. Permission from IEEE must be obtained for all
other uses, in any current or future media, including reprinting/republishing this material for advertising or
promotional purposes, creating new collective works, for resale or redistribution to servers or lists, or reuse
of any copyrighted component of this work in other works. 

\twocolumn

\maketitle
\begin{abstract}
Physical Unclonable Functions (PUFs) are gaining attention in the cryptography community because of the ability to efficiently harness the intrinsic variability in the manufacturing process. However, this means that they are noisy devices and require error correction mechanisms, e.g., by employing Fuzzy Extractors (FEs). Recent works demonstrated that applying FEs for error correction may enable new opportunities to break the PUFs if no countermeasures are taken. In this paper, we address an attack model on FEs hardware implementations and provide a solution for early identification of the timing Side-Channel Attack (SCA) vulnerabilities which can be exploited by physical fault injection. The significance of this work stems from the fact that FEs are an essential building block in the implementations of PUF-enabled devices. The information leaked through the timing side-channel during the error correction process can reveal the FE input data and thereby can endanger revealing secrets. Therefore, it is very important to identify the potential leakages early in the process during RTL design. Experimental results based on RTL analysis of several Bose–Chaudhuri–Hocquenghem (BCH) and Reed-Solomon decoders for PUF-enabled devices with FEs demonstrate the feasibility of the proposed methodology.\\

Keywords - timing side-channel attack, physical unclonable function, fuzzy extractor, fault-injection attack, error correction code, BCH, Reed-Solomon, RTL analysis.
\end{abstract}
\IEEEpeerreviewmaketitle

\section{Introduction}
Physical unclonable functions (PUFs) are hardware primitives which derive identifiers and cryptographic keys from the random variations of the silicon manufacturing process. PUFs provide a significantly higher security assurance as keys are volatile and derived only when required. Thus, a PUF can be easily attached or embedded into the cryptographic implementation for authentication and identification\cite{maes2010physically}.   
PUF-enabled devices are also an efficient alternative to the expensive conventional measures against the integrated circuit power-off, e.g., by using the Non-Volatile Memory (NVM) for the key storage. The keys generated by PUFs are derived by measurements in the field during the run time and can be saved in a cheaper volatile memory. 

PUFs are known to be sensitive to the environmental factors such as the ambient temperature, the supply voltage noise, etc. that may affect the reliability of the response measurement, and ultimately, reduce the reproducibility of the cryptographic key. Along with the external factors, the internal factors of the PUF’s manufacturing technology prevent it from guaranteeing a constant response all the time. This nondeterminism poses issues for applying a PUF as a key generator or identifier \cite{maes2009soft}. Therefore, for the post-processing, a Fuzzy Extractor (FE) is an essential component to help a PUF generate a reliable key by correcting the errors caused internally or by environmental variations.

 Different types of the PUF structure and the environmental conditions imply different requirements for the FE and the corresponding ECC. An example of a silicon PUF is the memory-based PUF, which is widely used in chip-level authentication. FE ECCs such as the Bose–Chaudhuri–Hocquenghem (BCH) \cite{maes2009soft} or Reed-Solomon\cite{korenda2019proof} are used in memory-based PUF enabled devices. 

While FEs with ECCs significantly raise reliability, they can lead to new exploits such as allowing  an attacker to extract sensitive information by studying the behavior of ECC. Side-Channel Attacks (SCA) on ECC implementations have attracted particular attention of the research community. In \cite{2017EM_side}, the authors extract the information about the key by non-invasive measurement of electromagnetic radiation together with a differential power analysis of the BCH decoder. In \cite{power_side_channel2010differential},
the authors study the simple power analysis of both BCH and Reed-Solomon code and manage to recover the PUF response from the collected power traces. 
However, there is no research work that refers to attacks that combine timing SCA and fault attacks for FEs, namely targeting to the execution time of the error-correcting code of FE in combination with the insertion of faults to PUF. So in this paper, we address this gap by a study on BCH and Reed-Solomon RTL designs execution time differences as a reaction to intentionally triggered faults inserted to PUF. Specifically, the contributions of the paper include:
\begin{itemize}
    \item Definition of an attack model based on fault injection and timing analysis of ECC execution that may lead to the secret PUF values extraction. 
    \item An early design stage RTL methodology for verification of an ECC design invulnerability against the proposed attack by employing both structural and simulation-based analysis steps. 
    \item Case studies of Reed-Solomon and BCH based ECC with vulnerabilities identification and exploitation.
\end{itemize}

The rest of the paper is organized as follows. Section II reviews the background of the FE architecture and ECC decoders. The attack model is discussed in Section III.  Section IV presents the proposed methodology for verifying invulnerability against the proposed attack. Section V presents a case study for ECC implementations. Section VI concludes the paper. 

\section{Background And Related Works}
\subsection{Fuzzy Extractor and Secure Sketch } 
\begin{figure}[h]
        \centering
        \includegraphics[scale=0.5]{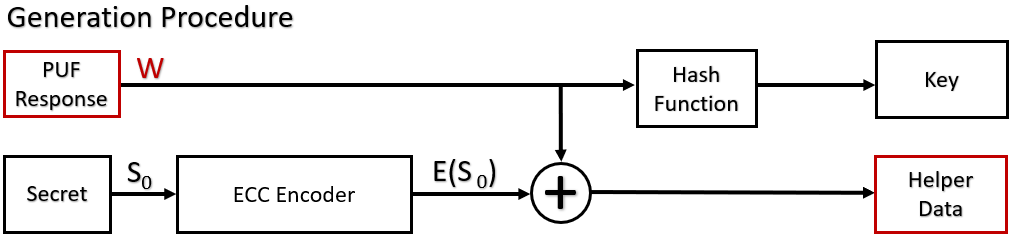}
        \caption{Generation Procedure in A PUF Fuzzy Extractor}
        \label{fig:fes_gen}
    \end{figure}

\begin{figure}[h]
        \centering
        \includegraphics[scale=0.5]{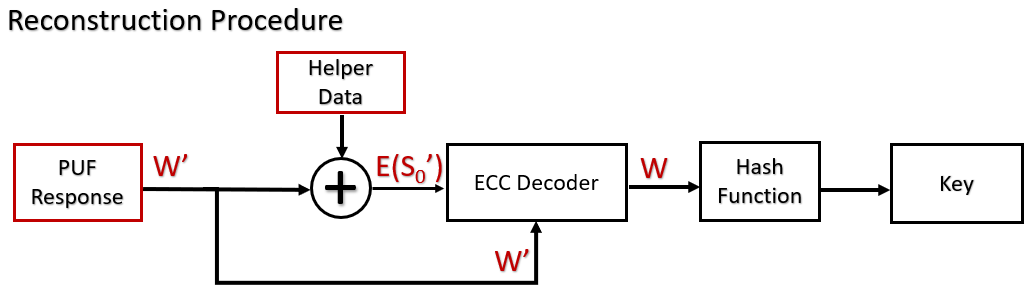}
        \caption{Reconstruction Procedure in A PUF Fuzzy Extractor}
        \label{fig:fes_rep}
    \end{figure}

The Fuzzy Extractor
\cite{dodis2004fuzzy}
is a secure method to generate cryptographic keys from noisy sources. The FE serves as a post-processing unit in memory-based PUF-enabled cryptographic schemes. It is used both in the Generation and Reconstruction Procedures, as illustrated in Fig. \ref{fig:fes_gen} and Fig. \ref{fig:fes_rep} correspondingly. 

In the Generation Procedure case, the fuzzy data from the PUF response $W$ and a random secret $S$ are used to generate the Helper Data by XOR operation on $W$ and $E(S_{0})$ which is encoded $S_{0}$. The generated helper data is stored in a non-volatile memory. In memory-based PUF-enabled devices, the Generation Procedure happens only once at the first-time power-on of the memory-based PUF.


On the contrary to this, the Reconstruction Procedure is executed many times during the product lifetime. Due to the noise and PUF manufacturing
randomness, it is difficult to generate the same response consistently. To reproduce the correct cryptographic key, the Helper Data, stored in an NVM, is used in conjunction with the measured PUF response $W\textprime$. Then with the help of the ECC decoder to detect and correct the divergent bits, the correct $W$  is reproduced. After applying the Hash Function, the expected correct cryptographic key is reconstructed.\\

The FE guarantees that the resulting key is consistent while the publicly accessible Helper Data does not leak any information related to the secret of the key. To ensure consistent generation of the correct key, the hamming distance between the measured PUF response $W'$ with the originally measured W in the Generator Procedure should be smaller or equal to the correction capability of the ECC decoder, represented as a constant value \textbf{\textit{t}}. In this paper, we assume that the measured responses of the memory-based PUF are within this hamming distance constraint.

Recent research works have identified potential attacks on FEs \cite{merli2011side}. Most of them target the Reconstruction Procedure.
In \cite{becker2017robust}, the authors report on  a method to extract the PUF secret by manipulating the Helper Data in the Reconstruction Procedure. In \cite{delvaux2014helper}, Delvaux et al. provide an in-depth analysis of the Helper Data algorithms, and identify new threats for leaking the Helper Data and the soft-decision coding. 

\subsection{ECC decoder}
The ECC unit is the main component in a FE. Binary BCH and Reed-Solomon are the two types of ECC that are widely used in PUF-enabled devices. 
Both codes are cyclic and capable of detecting up to $2t$ and correct up to $t$ errors by adding $2t$ check bits or non-binary values (symbols) to the data. Binary BCH is used for binary error correction, and Reed-Solomon is used for symbol error correction. While both software and hardware implementations of these codes exist, the hardware ones are more adopted. First, this is because the complex algorithms of the decoders require significant computational power along with the real-time constraints.  The second difficulty for software implementations is the limited support of the Galois Fields Arithmetic operation in the general-purpose processors \cite{riley2003introduction}. 
The hardware implementations of binary BCH and Reed-Solomon decoders are discussed in more detail in Section V.

\section{Attack Model}
In this paper, we assume an attack combining 1) fault injection to the memory-based PUF with 2) a timing SCA for observing and comparing the different decoding execution times of the ECC unit that is aimed at revealing the correct memory-based PUF data. In case of success, the attack explicitly compromises the core function of the PUF-enabled cryptographic devices, because the attacker can clone the PUF and can steal the secret. 


\subsection{Fault Injection Parameters}
For the physical fault injection to the memory-based PUF the following fault parameters are assumed.
\begin{enumerate}[label=(\alph*)]
    \item Granularity: each fault injection results in exactly one fault in one-bit data. 
    \item Modification (fault type): after the fault injection, the manipulated data is set to a specified logic value, i.e. either ’1’ or ’0’.
    \item Control: the attacker has a bit-wise precise control of fault injection to the memory-based PUF bits.
    \item Effect of the fault: the injected faults have a transient nature, i.e. the injected values are overwritten by the normal functionality of the device (e.g. the next measurement of the PUF on power-on). 
\end{enumerate}
Several studies on laser fault injection \cite{roscian2013fault} have demonstrated similar attack parameters and, therefore, the feasibility of the above assumptions. Technical details of the fault injection attack implementation are out of the scope of this study.

\subsection{Attack Assumptions}
The following set of assumptions must be satisfied for the success of the attack. The feasibility of the assumptions (iii)-(vi) is supported by several research works in state of the art.
\begin{enumerate}[label=(\roman*)]
    \item The output of a memory-based PUF measurement in the cryptographic device is processed by a FE with a binary BCH or Reed-Solomon based ECC. 
    
    \item The ECC implementation leaks exploitable information through the timing-side channel. 
    
    \emph{Comment:} The methodology for identifying the vulnerability enabling this assumption is the core contribution of this paper and presented in Section IV. 
    
    \item The memory-based PUF is noise-free under stable environmental conditions. The errors in the memory-based PUF are caused by the environment.
    
    \emph{Comment:} While an ideal noise-free memory-based PUF would not require the FE at all, we assume that the noise is caused by the variations in the external environment while the internal noise is negligible. 
    \cite{gao2019building} demonstrated that the external environmental conditions like the ambient temperature, supply voltage, etc. have a significant impact on the error rate of the PUF.
    
    \item The generated Helper Data is stored in NVM or the flash memory of the cryptographic devices and remains constant during the Reconstruction Procedure.
    
    \emph{Comment:} 
    As an added value, this assumption creates an advantage for the proposed attack, compared to alternatives (e.g. \cite{becker2017robust, delvaux2014helper}), because it does not rely on the attacker being able to modify the Helper Data.
    
    \item The fault injection parameters (a) to (d) hold (see III.A).
    
    \emph{Comment:} Several research works proposed bit-wise fault injection in SRAM and other on-chip memories. E.g., in \cite{skorobogatov2002optical}, bit-wise faults were successfully injected in a PIC microcontroller through a semi-invasive method and without mechanical damage to the silicon.
    
    \item The attacker has a controlled access for measuring the decoding execution time. 
    
    \emph{Comment:} The physical measurement of the ECC decoding execution time can exploit the reflection of timing by the power traces. In \cite{krieg2011side}, the authors analyze use of the AES execution power traces for a SCA. The power traces are represented by changes of power over time, with the timing information embedded. A similar approach is used in\cite{lai2019pascal} for RTL verification of RSA designs against vulnerability to timing SCAs. 
    
\end{enumerate}

\subsection{Attack Procedure}
\begin{figure}[!h]
        \centering
        \includegraphics[scale=0.45]{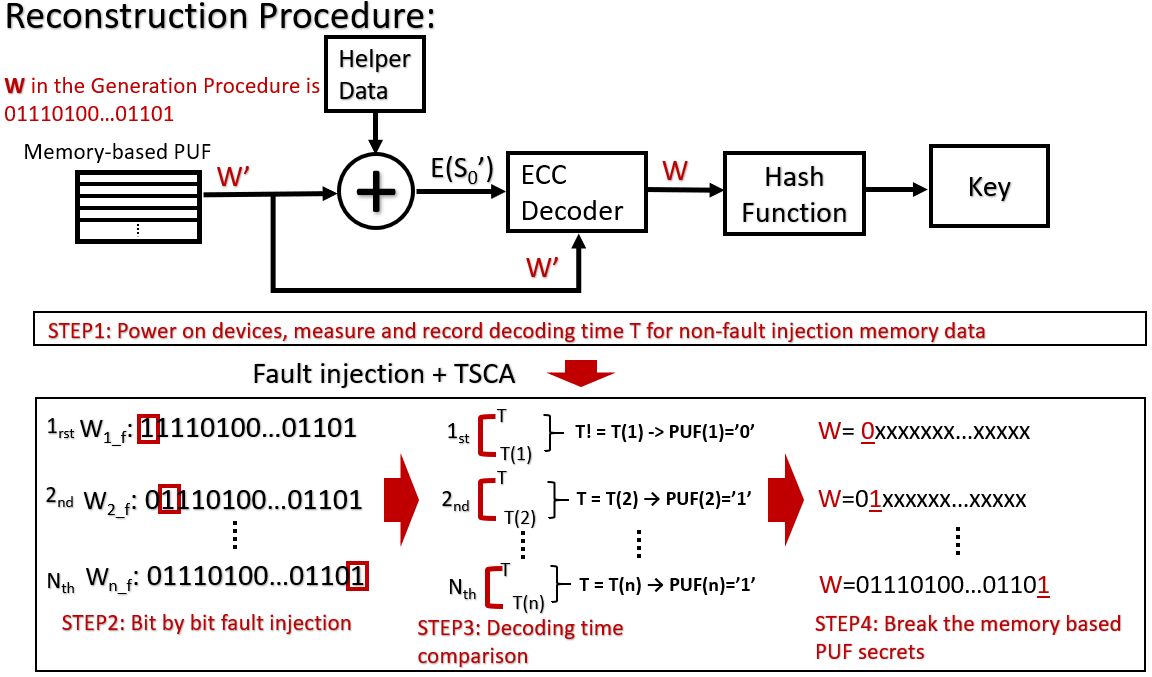}
        \caption{An illustration of the proposed attack procedure}
        \label{fig:f_i}
    \end{figure}

The proposed attack is a combination of fault injection with timing side channel analysis and represented by the following 4 steps. The procedure is illustrated in Fig.\ref{fig:f_i}. 
\begin{enumerate}
   \item Power on the device. 
   Measure the initial PUF data. With the above assumptions, this memory value should be error-free, i.e. the same with $W$ generated in the Generation Procedure. Measure and record the reference time $T$ as the number of clock cycles for the execution of the ECC decoding.
   \item Inject a fault $f$ at the $m_{th}$ bit of memory-based PUF following the (a) to (d) parameters and generate the new memory data $W_{m\_f}$. $W_{m\_f}$ has a one-bit difference value compared to $W$. E.g, if the $f$ is a set to logic “1” value and $m=1$ then $W$ and $W_{1\_f}$ can be either equal or can be different by exactly one bit at the first position. Then execute the Reconstruction Procedure, measure the decoding execution time $T(m)$.
   \item The relation between these two decoding times $T$ and $T(m)$ contains only two possible cases. The PUF's secret single bit $m$ can be revealed by comparing the two decoding times as follows:
        \begin{itemize}
             \item if $T != T(m)$, then a different value at the $m_{th}$ bit was injected. E.g., for $f=1$, the original value of the $m_{th}$ bit in memory is ’0’;
             \item if $T = T(m)$ then the value at the $m_{th}$ bit was equal to the injected one. E.g., for $f=1$, the original value of the $m_{th}$ bit in memory is ’1’;
        \end{itemize}  
    \item Repeat the steps 1) to 3). 
    of the procedure until the last $m_{th}$ bit of memory-based PUF. The memory-based PUF's secret value is revealed.
\end{enumerate}

\section{Proposed Methodology}
The precondition for the introduced attack is the non-constant decoding execution time in case of different input data for the ECC unit of the memory-based PUF Fuzzy Extractor. In this section, we propose a methodology to identify this vulnerability in an ECC implementation already at the RTL design phase. The methodology employs both structural and simulation-based analysis for binary BCH and Reed-Solomon algorithms based hardware ECC implementations. In practice, these two algorithms are widely used by the industry in memory-based PUF-enabled devices. 
\subsection{Structural Analysis of ECC Decoder}
\subsubsection{Binary BCH Decoder}
\begin{figure}[!ht]
        \centering
        \includegraphics[scale=0.45]{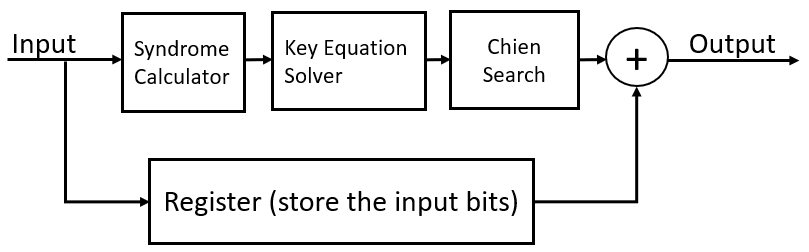}
        \caption{Binary BCH Decoder Structure}
        \label{fig:bch}
    \end{figure}

A general binary BCH decoder hardware implementation has three stages, as shown in  Fig.\ref{fig:bch}. The divergent (error) bits are identified by the Syndrome Calculator, Key Equation Solver and the Chien Search. Next, the decoder corrects the error bits by the XOR operation on the stored input with the identified error bits to recover the correct codeword. Let r(x), c(x) and e(x) be the received polynomial, codeword polynomial and error polynomial, i.e. r(x) = c(x) + e(x). Assume the binary BCH decoder can correct t errors.
As the structural analysis of the binary BCH, we consider the following reasoning. 
\begin{itemize}
\item \textbf{Syndrome Calculator:} It is the first stage in the decoder generates 2t syndromes as defined in (\ref{eq:syndrome}).
\begin{equation}
    S_{i}=r(x^{i})= r_{0}+r_{1}x^{i}+r_{2}x^{2i}+..... +r_{n-1}x^{(n-1)i}  
    \label{eq:syndrome}
\end{equation}
where $1\leq i \leq 2t-1$.
An important feature of the syndromes is that they do not depend on transmitted information but only on error locations. If at position $i$ there is an error then $S_{i}$ has a non-zero value and it is equal to zero otherwise. For all possible inputs, the decoder always generates $2t$ syndromes. Therefore, the time for the syndrome calculation is constant for the BCH decoder with a fixed error correction capability.
\item \textbf{Key Equation Solver:} In the second stage, the error location polynomial  $\sigma (x)$ is generated. Berlekamp Massey Algorithm (BMA) is one known iterative procedure that determines polynomial equation (\ref{eq:bma}) out of a set of linear equations for the 2t syndromes calculated in the first stage.
\begin{equation}
     \sigma (x) = 1+\sigma_{1}x+\sigma_{2}x^{2}+...\sigma_{t}x^{t}   
      \label{eq:bma}
\end{equation}
    BMA can be implemented in parallel or serially. In \cite{liu2006low}, it is demonstrated that a parallel implementation for a t errors correction BMA needs 2t iterations. A serial implementation implies a significant increase in the number of iterations. According to \cite{chang1999new}, it needs $2t^{2}$ iterations. However, for both cases, the total number of iterations is determined only by t, which is the maximum number of errors the decoder can correct.
    \item \textbf{Chien Search:} This stage searches for error locations by checking the roots of  $\sigma(x)$. It is a simple trial-and-error procedure. All nonzero elements of the Galois Fields for a binary BCH decoder are generated in sequence and only capture the condition when $\sigma (x_{i})$ is equal to zero which the error position. Therefore, in this stage, the total number of nonzero elements depends only on the Galois Field GF($2^{m}$) where $n=2^{m}-1$ and n is the size of codeword. 
    \end{itemize}
To conclude, for different binary BCH decoder implementations, the error correction bits and the size of the codeword are the factors which lead to the different decoding execution time. However, for a specific binary BCH decoder, these parameters are fixed at the design phase. Therefore, the structural analysis has not identified timing channels in binary BCH decoder structures.

\subsubsection{Reed-Solomon Decoder}
\begin{figure}[!ht]
        \centering
        \includegraphics[scale=0.45]{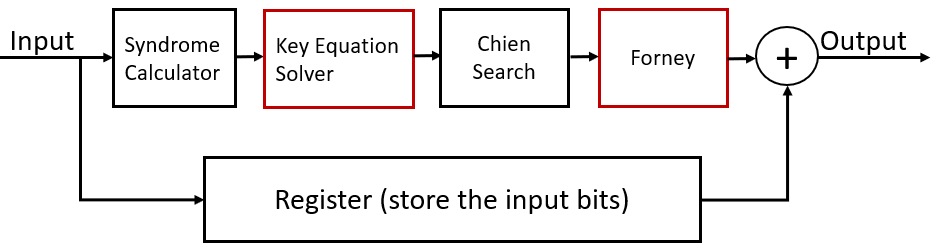}
        \caption{Reed-Solomon Decoder Structure}
        \label{fig:rs}
    \end{figure}

Reed-Solomon (RS) decoder 
aimes at non-binary (symbol) error correction. Different from the binary BCH, which needs only to generate error locator polynomial $\sigma (x)$ RS also needs to generate an error value polynomial. Therefore, some RS implementations replace BMA by Euclidean Algorithm (EA) for the Key Equation Solver to calculate the error location polynomial and error value polynomial and add a new component Forney to calculate the error value. The Reed-Solomon decoder structure is illustrated in Fig.\ref{fig:rs}. Here, the differences with the BCH decoder structure are highlighted in red. In the following structural analysis, we focus only on these two different components.

\begin{itemize}
    \item \textbf{Euclidean Algorithm (EA):} It is an iterative procedure to generate the \emph{error locator polynomial} and the \emph{error value polynomial} with the 2t syndromes generated by the Syndrome Calculator stage. Particular implementations of EA may prefer a pipelined version with the objective of performance optimization \cite{lee2008high}. In EA procedure \cite{lee2008high}, the error locator polynomial $\sigma(x)$ and the error value polynomial $\omega(x)$ are acquired by solving the equation (\ref{eq:omega}). Equation (\ref{eq:omega}) can be represented in the form of equation (\ref{eq:tr_0}). The extend Euclidean Algorithm can find a series polynomial by (\ref{eq:tr_1}). From (\ref{eq:tr_0}) and (\ref{eq:tr_1}), $A_{i}(x)=\sigma(x)$, $R_{i}(x)=\omega(x)$ and $B_{i}(x)=-Q(x)$. To solve the Key Equation the EA procedure starts with initiating the values $R_{0}(x)=x^{2t}$, $Q_{0}(x)=S(x)$, $L_{0}(x)=0$, $U_{0}(x)=1$ and then it is followed by interactions of four equations used to calculate $R_{i}(x)$, $Q_{i}(x)$, $L_{i}(x)$ and $U_{i}(x)$, based on the values from the previous stage, until the degree of $R_{i}(x)$ gets smaller than the degree of $L_{i}(x)$ or t. When the iteration is finished, the equation (\ref{eq:omega}) is solved. Because the $R(x)$ starts at the degree 2t, and the iteration can finish at the degree of $R(x)$ equal to t or smaller. Therefore, the EA stage may require a different number of iterations for the different codewords which may introduce different execution times.
    \begin{equation}
         \omega(x)=S(x)\sigma(x)\mod x^{2t}
         \label{eq:omega}
    \end{equation}
   \begin{equation}
       \sigma(x)S(x)= Q(x)x^{2t}+\omega(x)
       \label{eq:tr_0}
   \end{equation} 
   \begin{equation}
       A_{i}(x)S(x)+B_{i}(x)x^{t}=R_{i}(x)
   \label{eq:tr_1}
   \end{equation}
    
    \item \textbf{Forney:} By using the Forney algorithm, the error value e(x) can be acquired by the equation (\ref{eq:forney}).
    \begin{equation}
     e_{j}=-\dfrac{\omega(X_{j})}{\sigma^\textprime(X_{j})}
     \label{eq:forney}
    \end{equation}
  Normally, it is implemented in combinational logic because  $\omega(X)$ and $\sigma(x)$ are available. The execution time of this stage is constant.
\end{itemize}    
To conclude, the structural analysis has not identified the timing channel in the other stages of the Reed-Solomon structure but the second stage. Based on the implementation, the Key Equation Solver stage in the Reed-Solomon based ECC decoder can introduce the vulnerability.
\subsection{Simulation-based analysis of ECC decoder }

In an RTL simulation of an ECC decoder implementation, a number of stimuli data parameters may have an impact on the execution time of a decoding iteration. For the proposed simulation-based analysis step, the following parameters are identified:
\begin{itemize}
    \item $codeword_{value}$: the encoded codeword value
    \item $error_{value}$: the error value is relevant only for a non-binary (symbol) ECC decoders 
    \item $error_{position}$: the error bit position for a binary ECC decoder or the error symbol position for a non-binary ECC decoder 
    \item $error_{number}$ : the number of error bits or symbols for binary or non-binary ECC decoder correspondingly
\end{itemize}
The structural analysis of binary BCH and RS decoders and the defined attack model allows reducing the search space. Table \ref{tab:ecc_decoding} presents the relationship of the execution time variation introduced by manipulating a particular decoding parameter and the vulnerability to the proposed attack. The notations \emph{C} and \emph{NC} represent constant and non-constant decoding execution time, while \emph{V} and \emph{NV} represent vulnerability or invulnerability.

\begin{table}[h!]
\centering
\caption{ECC execution time variability and the SCA vulnerability}
\begin{tabular}{|c|c|c|}
\hline
 \multicolumn{3}{|c|}{ECC Decoding Execution Time/Vulnerability} \\\hline
Parameters  &  RS decoder & Binary BCH decoder\\\hline
$codeword_{value}$ &  C/NV & C/NV  \\\hline
$error_{value}$ &  C/V & \notableentry  \\\hline
$error_{position}$ & C/V & C/V  \\\hline
$error_{number}$ &  NC/V & C/V  \\\hline
\end{tabular}
\label{tab:ecc_decoding}
\end{table}
In particular, manipulation of the $codeword_{value}$parameter does not identify the vulnerability of the target decoder. The attacker does not have access to manipulate the predefined correct codeword and can only manipulate the input codeword to cause an error. Based on the structural analysis, it is already known that different codewords do not introduce different decoding time neither in binary BCH nor in RS structures. The $error_{value}$ and $error_{position}$ parameters can be manipulated by the attacker by injecting faults to the input codeword. However, the constant decoding time will not leak information through the timing channel. From Table \ref{tab:ecc_decoding}, we can conclude that the binary BCH decoder structures are secure with regards to the information leakage through the timing channel. An RS decoder implementation can be vulnerable if the attacker injects a different number of error symbols, i.e. the $error_{number}$. The table guides the designer which simulation campaigns are required to verify a particular implementation against vulnerability to the proposed SCA. 
\section{Case Study }
The feasibility of the proposed methodology was validated by running an exhaustive simulation campaign on 3 case study ECC designs for memory-based PUF Fuzzy Extractors, i.e. 2 binary BCH and a Reed-Solomon ECC implementations. 


\subsection{Binary BCH decoder}
The implementation of the binary BCH decoder is an open-source design in RTL Verilog accessible from Github\cite{GithubBCH}. Its general architecture is illustrated in Fig. \ref{fig:bch}. The decoder was configured for a 12-bit codeword, 8-bit message and supports two types of BMA, i.e. serial \emph{BMA\_serial} and parallel \emph{BMA\_parallel} versions. The configuration was set to correct up to two errors, i.e. $t=2$. Both versions were simulated with an exhaustive set of test vectors to identify the timing information leakage.
Only valid values for the 12-bit binary codeword were extracted by running the encoder with all possible inputs. The input for the encoder is 4-bit message and 2-bit error correction capability. Since the number of errors correctable for a given polynomial is sparse, the encoder has the selection algorithm to select suitable polynomial function to meet the provided requirements. Thus the actual message bit might be changed. In our case, the encoder pads 4-bit zeros and makes the input message bit 8-bit. We input all possible 4-bit value into encoder.
Then each encoded message value was merged with all possible error combinations considering the injection of 0, 1 or 2 errors at a time, i.e. all combinations of $error_{number}$ and $error_{position}$ were simulated. This means $T_{test\_vectors} = 2^{4} * (\binom{12}{0} + \binom{12}{1} + \binom{12}{2})$=1,264 ECC decoding executions were analyzed for the each design, and the decoding time was measured.
\subsection{Reed-Solomon decoder}
The case-study Reed-Solomon decoder implementation is also an open-source design accessible from Github \cite{GithubRS} and illustrated in Fig.\ref{fig:rs}. The design was configured for 8-symbol codewords, 4-symbol messages and 8-bit symbols. The error correction capacity was also set to 2 errors, i.e. $t=2$. By default, the design is pipelined by using registers to extend the execution time for each stage to the worst execution-time case. In practice, for memory-based PUF enabled devices where execution time is a critical factor, a configuration aimed at the decoder speed optimization is often used. This was also applied for the current case study. 
Different from the binary BCH, the Reed-Solomon decoder uses symbol-based error correction. While the parameter $error_{position}$ represents the position of the error symbol, the $error_{value}$ can take one of the $2^{8}=256$ possible values for an error in each symbol. The number of all combinations for the valid codewords merged with all possible errors for each symbol is $T_{test_vector} = \binom{8}{1}*(2^{8}-1) + \binom{8}{2}*(2^{8}-1) + \binom{8}{0} $= 1,822,741 that represents the number of executions to simulate and analyse per codeword. In the simulation campaign, we limited the analysis to one random valid codeword. Based on the architecture analysis, the other codewords provide the same results.

\subsection{Experiment Results Analysis}
Experiment results are shown in Table \ref{tab:tsca}. In the list of parameters identified for manipulation by the proposed methodology, the symbols "\solidcircle" and "-" represent the varied and constant parameters correspondingly. 
$T_{d}$ denotes the number of different decoding execution times identified and the corresponding values in clock cycles. 
For the Binary BCH, the experimental results confirm the conclusions of the structural analysis and do not identify any variations in the execution times. 
For the Reed-Solomon decoder, the red cells highlight the cases with the varying decoding time. In this experiment, $T_{d}$\emph{:3 \{38, 66, 72\}} denotes different timing cases in case of the different number of errors to be corrected, i.e. 
38, 66 or 72 clock cycles for 0, 1 or 2 errors correspondingly.  As shown in the first three rows, different $error_{position}$ and $error_{value}$ can not affect the decoding time, and it remains constant (but can be equal to different values) $T_{d}:1$ $\{38\}\|\{66\}\|\{72\}$. 

    \begin{table}[!h]
    \caption{ECC-based FE Decoding Timing Analysis}
    \resizebox{0.5\textwidth}{!}{
    \begin{tabular}{ |m{0.15cm}|m{0.15cm}|m{0.15cm}|m{0.15cm}|m{1.35cm}|m{1.4cm}|m{2.7cm}|}
    \hline
    \multicolumn{4}{|c|}{Varied Parameters} & \multicolumn{3}{|c|}{Decoding time by ECC Implementations (clock cycles)}\\ 
    \hline
    \rotatebox{90}{$codeword_{value}$} & \rotatebox{90}{$error_{number}$} &  \rotatebox{90}{$error_{position}$} &  \rotatebox{90}{$error_{value}$} &  Binary BCH-12-8 BMA\_serial & Binary BCH-12-8 BMA\_parallel & Reed-Solomon-4-8-8\\ 
    \hline
    - & - & - & \solidcircle & \notableentry & \notableentry & $T_{d}$:$1\,\{38\}\|\{66\}\|\{72\}$  \\ 
    \hline
     - & - & \solidcircle & - & $T_{d}$:$1\,\{28\}$ & $T_{d}$:$1\,\{21\}$ & $T_{d}$:$1\,\{38\}\|\{66\}\|\{72\}$ \\ 
     \hline
     - & - & \solidcircle & \solidcircle & \notableentry & \notableentry & $T_{d}$:$1\,\{38\}\|\{66\}\|\{72\}$  \\ 
     \hline
     - & \solidcircle & - & - & $T_{d}$:$1\,\{28\}$ & $T_{d}$:$1\,\{21\}$ & \cellcolor{red!20} $T_{d}$:$3\, \{38, 66, 72\}$\\ 
     \hline
     - & \solidcircle & \solidcircle & - & $T_{d}$:$1\, \{28\}$ & $T_{d}$:$1\, \{21\}$ & \cellcolor{red!20}$T_{d}$:$3\,\{38, 66, 72\}$ \\ 
     \hline
     - & \solidcircle & - & \solidcircle & \notableentry & \notableentry & \cellcolor{red!20}$T_{d}$:$3\, \{38, 66, 72\}$ \\ 
     \hline
     - & \solidcircle & \solidcircle & \solidcircle & \notableentry & \notableentry &\cellcolor{red!20} $T_{d}$:$3\, \{38, 66, 72\}$ \\ 
     \hline
     \solidcircle & - & - & - & $T_{d}$:$1\, \{28\}$ & $T_{d}$:$1\, \{21\}$ & \notableentry  \\ 
     \hline
      \solidcircle & \solidcircle & - & - &$T_{d}$:$1\, \{28\}$ & $T_{d}$:$1 \,\{21\}$ & \notableentry\\ 
     \hline
     \solidcircle & \solidcircle & \solidcircle & - &$T_{d}$:$1\, \{28\}$ & $T_{d}$:$1\, \{21\}$ & \notableentry\\ 
     \hline
      \solidcircle & - &  \solidcircle & - & $T_{d}$:$1\, \{28\}$ & $T_{d}$:$1\, \{21\}$ & \notableentry \\ 
     \hline
    \end{tabular}
    }
    \label{tab:tsca}
    \end{table}
\section{Conclusions}
Application of Fuzzy Extractors for error correction may enable opportunities to break the secure PUFs if no countermeasures are taken. This paper considers a combined attack model based on fault injection and timing analysis of ECC execution. In the worst case, such an attack may lead to the secret PUF value extraction. An early design stage RTL methodology was developed to verify the ECC design invulnerability against such or a similar SCA. 

The methodology involves structural and simulation-based analysis parts. In our study, we targeted at two ECC architectures most widely used in FEs. The structural analysis has not identified vulnerabilities in the considered binary BCH architectures, while the architecture of Reed-Solomon based ECC may be vulnerable in particular implementations. A set of simulation-based experimental results have confirmed the findings and demonstrated the timing information leakage. Under the specified assumptions, the proposed attack procedure is able to exploit this vulnerability and reveal the secret.

The results of the early RTL analysis can guide in the selection of suitable ECC implementation or in the application of design-level countermeasures. To remove the leakage, e.g., a register can be added at the output of the Euclidean Algorithm stage to equalize the timing to the worst-case execution, or optimizations at the ECC algorithm may be applied. The efficiency of the mitigation solutions can be explored by the proposed methodology at a low cost.


\section{acknowledgements}
\par {\small This research was supported in part by the project H2020 MSCA ITN RESCUE funded from the EU H2020 programme under the MSC grant agreement No.722325 and by European Union through the European Structural, Regional Development and Social Funds.}
\bibliographystyle{./bibliography/IEEEtran}
\bibliography{./bibliography/IEEEabrv,./bibliography/IEEEexample}

\end{document}